\documentstyle[preprint,prl,tighten,aps,epsfig]{revtex}

\begin{document}
\def\slash#1{#1\!\!\!\!\!/}
\def\rpv{\slash{R_p}~}

\draft

\preprint{ 
LANCS--TH/0005\hspace{1ex} 
KAIST--TH 00/07 \hspace{1ex} 
KIAS--P00018}

\title{$CP$ Asymmetries in Radiative B Decays with R--parity Violation}

\author{Eung Jin Chun$^{a,b}$, Kyuwan Hwang$^c$ and Jae Sik Lee$^b$}

\address{
\vspace{0.2 cm}
$^a$Department of Physics, Lancaster University, Lancaster LA1 4YB, UK\\
$^b$Korea Institute for Advanced Study, Seoul 130--012, Korea\\
$^c$Department of Physics, KAIST, Taejon 305-701, Korea}

\maketitle

\begin{abstract}
We analyze the effect of R--parity violation in the minimal supersymmetric
standard model on the $CP$ asymmetries 
in $b\to s \gamma$ decay.
The direct and mixing-induced $CP$ asymmetries arising from the lepton 
number violating couplings are strongly constrained by the current 
experimental limits on the corresponding couplings.  
Allowing a heavy neutrino ($m_{\nu_\tau}\sim10$ keV) and a moderate 
mass splitting of sfermions,  the direct $CP$ asymmetry around 15 \% 
and the nearly maximal mixing-induced $CP$ asymmetry ($\sim 100 \%$) 
can be realized, depending on the R--parity conserving contributions 
to the radiative $b$ decay. 
With the baryon number violating couplings, only the mixing-induced 
$CP$ asymmetry arises and it can be maximal 
provided there is a the similar sfermion mass splitting.
\end{abstract}



\section{Introduction}
\label{sec:introduction}
$CP$ violation in radiative $B$ decays may provide a promising tool for
probing new physics beyond the Standard Model (SM).
The direct $CP$ asymmetry  in the radiative $b\to s\gamma$ decay defined by
\cite{kagan}
\begin{equation} \label{ACP}
A_{CP}^{b\rightarrow s\gamma}(\delta)=\left.
\frac{
\Gamma(\overline{B}\rightarrow X_s\gamma)-
\Gamma(B\rightarrow X_{\overline{s}}\gamma) }{
\Gamma(\overline{B}\rightarrow X_s\gamma)+
\Gamma(B\rightarrow X_{\overline{s}}\gamma) }
\right|_{E_\gamma > (1-\delta)E_\gamma^{\rm max}}\,
\end{equation}
is below 1 $\%$ in the SM \cite{soar}.
Therefore, the observation of a sizable $CP$ asymmetry would be a clean
signal of new physics and may further discriminate various extensions
of the SM.  Recent model-independent analyses of $CP$ violating
effects in the inclusive decays $B\to X_s \gamma$  in terms of
the Wilson coefficients of the dipole moments operators have shown
that models with enhanced chromo-magnetic dipole moments can naturally 
provide a large asymmetry \cite{kagan} and it  can reach up to 
30 $\%$ accommodating the observed branching ratios of $b\to s\gamma$
and $b\to sg$ decays \cite{ko}. The specific predictions for the
$CP$ asymmetry are of course model-dependent.  The left-right model 
can yield $A_{CP}$ at the level of 1 $\%$ \cite{asat},
and a two Higgs doublet model up to 10 $\%$ \cite{wolf}.
In supersymmetric extensions of the Standard Model, there exists additional
$CP$ phases which would give rise to large $CP$ violating effects.
In the context of minimal supergravity models \cite{nilles}, 
the direct $CP$ asymmetry turns out to be less than 
2 $\%$ due to strong constraints on supersymmetric $CP$ violating phases 
from the neutron and electron electric dipole moments \cite{keum}.
The asymmetry can be enlarged in non-minimal models up to 7 $\%$ with
more freedom in $CP$ phases \cite{cho}, or up to 15 $\%$ with generic 
sfermion mixing \cite{chua,ko}.

\medskip

Another important $CP$ violating observable is the  mixing-induced $CP$
asymmetry in exclusive radiative $B$ decays \cite{atwood} which can 
occur for radiative $B_q \to M_q \gamma$ decays, where  $M_{q=d,s}$ 
is any hadronic self-conjugate state with $CP$ eigenvalue $\xi=\pm 1$.
The $CP$ asymmetry in the time-dependent decay rates $\Gamma(t)$ 
for $B_q\to M_q\gamma$ and $\bar{\Gamma}(t)$ for $\bar{B_q} \to M_q \gamma$
is then
\begin{equation} \label{AM}
 A(t)\equiv {\Gamma(t)-\bar{\Gamma}(t) \over \Gamma(t)+\bar{\Gamma}(t)}
 = \xi A_M \sin(\phi_M-\phi_L-\phi_R)\sin(\Delta m t) \,, \quad
 A_M \equiv {2 |C_{7L}C_{7R}| \over |C_{7L}|^2+|C_{7R}|^2}
\end{equation}
where $\phi_M$ and $\Delta m$ are  the phase and the mass difference 
of $B_q$--$\bar{B}_q$ mixing, respectively, and  
$C_{7L,7R}$ are the effective coefficients of the left-handed and 
right-handed dipole moment operators for the $b\to q \gamma$ decays,
and $\phi_{L,R}$ are their phases, respectively.  
In the standard model, such asymmetries are expected to be small 
as one has $C_{7R}/C_{7L} \approx
m_q/m_b$ and thus $A_M$ of order 1 $\%$ and 10 $\%$ for the
$b\to d\gamma$ and $b\to s\gamma$ decays, respectively.
Unlike the direct $CP$ asymmetry, the left-right symmetric model may allow
the mixing-induced asymmetry up to 50 $\%$ \cite{atwood}.  The supersymmetric
model with generic sfermion mixing can yield even larger asymmetry
up to 90 $\%$ \cite{chua}.

\medskip

In this paper, we will analyze the effects of R--parity violation in the 
Minimal Supersymmetric Standard Model (MSSM) on the $CP$ asymmetries in 
radiative $B$ decays.   
%
%
R--parity violation in the MSSM introduces a large number of 
trilinear couplings which violate lepton and baryon number.
These additional couplings can surely be  sources of flavor and $CP$
violation, which might lead to a huge effect on the $CP$-odd observables
in the B decays.  
$CP$ violating effects of R--parity violation 
in the hadronic $B$ decays have been considered previously
in Refs.\cite{worah,kaplan,guetta,jang,datta}, showing that
significant modifications to the SM predictions can follow from
R--parity violation.
This work is devoted to the analysis of the $CP$ asymmetries
in the radiative $B$ decay.
As we will see, contrary to the cases with hadronic $B$ decays,
various experimental constraints on the R--parity violating couplings 
coming particularly from rare $B$ decays strongly limit the 
amount of the direct $CP$ asymmetry in the $b\to s\gamma$ decay
whereas a large mixing-induced $CP$ asymmetry can be induced from,
in particular, baryon number violation.  

This paper is organized as follows.  In section II, we analyze 
the general effective Hamiltonian describing the $b\to s\gamma$ decay
including the new  operators induced by R--parity violation.
We derive the anomalous dimensions and evolution matrix with the
enlarged operator set.
In section III, we discuss the $CP$ asymmetries arising from
lepton number violating couplings with which we can have both the
direct and mixing-induced $CP$ asymmetry. We deal with both cases 
in separated subsections.
In section IV, we discuss the $CP$ asymmetries arising from
baryon number violating couplings, in which case only 
mixing-induced $CP$ asymmetry can be obtained.
We conclude in section V.


\section{R--parity violation and effective Hamiltonian}

Let us begin our discussion with defining our choice of the basis
to describe the R--parity violating couplings.
For comparison with experiments, it is convenient to work with R--parity 
violating couplings defined in the quark and lepton mass eigenbasis.  
In this prescription, we leave the neutrinos in the charged lepton mass 
eigenbasis since 
neutrinos can be taken to be massless for our purpose.  
The full superpotential of the MSSM fields including generic
R--parity violating  couplings is then given by
\begin{eqnarray}  \label{supo}
W &=& \mu H_1 H_2 + h^e_i H_1 L_i E_i^c + 
      h^d_i (H_1^0 D_i D_i^c - H_1^- V^\dagger_{ij} U_j D^c_i)
      + h^u_i(H_2^0 U_i U_i^c -H_2^+ V_{ij} D_j U_i^c) \nonumber \\
&+&  {\lambda}_{ijk}(L^0_i E_j-L_j^0 E_i) E_k^c 
+ {\lambda}'_{ijk} (L_i^0 D_j D_k^c -  E_i V^\dagger_{jl} U_l D^c_k) 
+ {1\over2}\lambda''_{ijk} U^c_i D^c_j D^c_k \,,
\label{WZ}
\end{eqnarray} 
where $L_i=(L_i^0, E_i)$ and $Q_i=(U_i, D_i)$ are the lepton and 
quark $SU(2)$ doublets,
and $E_i^c$, $U_i^c$, $D_i^c$ are the $SU(2)$ singlet
anti-lepton and anti-quark superfields.
Here $V_{ij}$ is the Cabibbo--Kobayashi--Maskawa (CKM) matrix of quark fields.
In order to ensure the longevity of  proton, the  products  
$\lambda' \lambda''$ have to be highly suppressed  \cite{ptri}.
For this reason, one usually assumes lepton or baryon number conservation
to discard the couplings $\lambda/\lambda'$ or $\lambda''$, respectively.
In this paper, we discuss both cases separately: 
one with R--parity and lepton 
number violation with nonvanishing $\lambda'$, and the other
with R--parity and baryon number violation with $\lambda''$.
The couplings $\lambda$ will be irrelevant for our discussion.

\medskip

The presence of R--parity violating couplings in Eq.~(\ref{supo})
give rise to new contributions to the radiative decay $b \to s \gamma$  
through one-loop diagrams  exchanging sleptons or squarks 
\cite{carlos} as depicted in FIG.~1.  
Note that R--parity violation may induce equally sizable dipole moments 
of the left-handed and right-handed type, respectively  labeled by $L$ 
and $R$ as follows:
\begin{equation} \label{O7}
O_{7L}\propto\overline{s^\alpha_L}\sigma^{\mu\nu} 
b^\alpha_R F_{\mu\nu} \,, \quad 
O_{7R}\propto\overline{s^\alpha_R}\sigma^{\mu\nu} 
b^\alpha_L F_{\mu\nu} \,.
\end{equation}
In the SM and many extensions of it, the coefficients of the
second one is usually suppressed by the factor $m_s/m_b$.
In our case with lepton or baryon number violation in the MSSM,
the operator $O_{7L}$ and $O_{7R}$ are generated for nonvanishing
combinations of couplings;
\begin{equation} \label{lamLR}
\lambda'_{n3j}\lambda'^*_{n2j} \quad {\rm and} \quad
\lambda'_{nj2}\lambda'^*_{nj3}/\lambda''_{n12}\lambda''^*_{n23} \,,
\end{equation}
respectively.  
With the couplings in Eq.~(\ref{lamLR}), there arise also other
four-quark operators which should be taken into account in 
the complete effective Hamiltonian describing the $b\to s \gamma$ decay.
The whole set of the effective operators arising from
R--parity violation can be described by a simple generalization
of the SM operator space by separating the standard
operators $O_{3,4,5,6}$ for each quark flavor $q=u,c,d,s,b$.
That is, we introduce the additional operators, 
\begin{eqnarray} \label{OL}
O^{q}_{3L}=\overline{s^\alpha_L}\gamma^\mu b^\alpha_L \,
\overline{q^\beta_L}\gamma_\mu q^\beta_L \,, 
&\quad&
O^{q}_{4L}=\overline{s^\beta_L}\gamma^\mu b^\alpha_L \,
\overline{q^\alpha_L}\gamma_\mu q^\beta_L \,, \nonumber \\
O^{q}_{5L}=\overline{s^\alpha_L}\gamma^\mu b^\alpha_L \,
\overline{q^\beta_R}\gamma_\mu q^\beta_R \,, 
&\quad&
O^{q}_{6L}=\overline{s^\beta_L}\gamma^\mu b^\alpha_L \,
\overline{q^\alpha_R}\gamma_\mu q^\beta_R \,, 
\end{eqnarray}
where $\alpha,\beta$ are the color indices.
The dipole moment operators for the $b \to s\gamma$ and $b\to s g$ 
are defined by
\begin{eqnarray} \label{78L}
O_{7L}&=&\frac{e}{16\pi^2}m_b\overline{s^\alpha_L}\sigma^{\mu\nu}
b^\alpha_R F_{\mu\nu} \,, \nonumber \\
O_{8L}&=&\frac{g_s}{16\pi^2}m_b\overline{s^\alpha_R}\sigma^{\mu\nu}
T^a_{\alpha\beta} b^\beta_L G^a_{\mu\nu} \,.
\end{eqnarray}
We also have the right-handed counterpart of the operators
which can be obtained by the exchange $L \leftrightarrow R$ in 
Eqs.~(\ref{OL}) and (\ref{78L}).  

The effective Hamiltonian at scale $\mu\leq {\cal O}(m_W)$ 
relevant for the $b\to s \gamma$ decay is now described in terms of
the enlarged set of operators as follows:
\begin{eqnarray} \label{Heff}
{\cal H}_{\rm eff}&=&
-\frac{4G_F}{\sqrt{2}}\lambda_t
\left[\sum_{i=1}^8C_{iL}(\mu)\,O_{iL}(\mu)
+\sum_{q}\sum_{j=3}^6 C^{q}_{jL}(\mu)\,O^{q}_{jL}(\mu) + 
(L \leftrightarrow R) \right]\,,
\end{eqnarray}
where $\lambda_t=V^*_{ts}V_{tb}$, $G_F$ is the Fermi constant
and $C$'s are the Wilson coefficients
which will be determined later.  The operators $O_{iL}$ with 
$i=1,\cdots,8$ are those considered in the SM \cite{buras}.
Note that $O_{1L,2L}=  O^c_{3L,4L}$ and  
$O_{iL}=\sum_q O^q_{iL}$ for $i=3,\cdots6$.

\medskip

Given the Wilson coefficients $C_i, C_j^q$ including the contributions
from R--parity violation at the weak scale $\mu=m_W$, 
we need to calculate those at the scale 
$\mu \sim m_b$ through the renormalization group (RG) evolution.  
Since the QCD running do not mix the left-handed and right-handed set 
of operators and its effect is identical, it is enough to calculate 
the RG equation at the one sector.  
At the leading order, it is rather straightforward to calculate 
the anomalous dimension matrix in the extended operator basis
following the standard calculation \cite{buras}.  
At this point, let us recall that it is convenient to use the so-called
``effective coefficients'' \cite{bmmp} which are free from the 
regularization-scheme dependency in the mixing between the sets
$O_i^{(q)}$ with $i=1,\cdots,6$ and $O_{7,8}$ resulting from two-loop
diagrams.   In terms of the effective coefficients, 
\begin{eqnarray} \label{Ceff}
C^{eff}_7(\mu)&=& C_{7}(\mu)+ \sum_I y_I C_I(\mu) 
\nonumber\\
C^{eff}_8(\mu)&=& C_{8}(\mu)+ \sum_I z_I C_I(\mu) \,,
\end{eqnarray}
with the index $I$ running for 26 indices labeled by $i$ and $jq$ for the
operators $O_i$ ($i\neq 7,8$) and $O^q_j$, 
the effective anomalous dimension matrix is given by
\begin{equation}
\gamma^{eff}_{JI} = \cases{ 
\gamma_{J7} + \sum_K y_K \gamma_{JK} - y_J \gamma_{77} -z_J \gamma_{87}\,,
\quad\mbox{for $I=7, J\neq 7,8$}  \cr
\gamma_{J8} + \sum_K z_K \gamma_{JK}  -z_J \gamma_{88}\,,
\quad\mbox{for $I=8, J\neq 7,8$}  \cr
\gamma_{JI} \,,
\quad\mbox{otherwise.} \cr }
\end{equation}
The corresponding RG equations for the effective coefficients are then
\begin{equation}
{d\over d\ln\!\mu} C^{eff}_I(\mu)= 
{\alpha_s \over 4\pi} \gamma^{eff}_{JI} C^{eff}_J(\mu) \,.
\end{equation}
In the naive dimensional regularization scheme which we follow \cite{bmmp},
the nonvanishing coefficients $y_I, z_I$ are 
\begin{eqnarray} \label{yyz}
&&y_5=y_{5b}=-{1\over3}\,,\quad y_6=y_{6b}=-1 \nonumber\\
&&z_5=z_{5b}=-1  \,.
\end{eqnarray}
Now, keeping track of the flavor structure of the standard calculation 
of the matrix $\gamma^{eff}$ \cite{buras}, one can 
find the $28\times 28$ anomalous dimension matrix 
in a straightforward way.  The results are presented 
in the Appendix.  There, we also calculate  the evolution matrix
from which the Wilson coefficients at the scale $\mu=m_b$ can be 
obtained in terms of the coefficients determined at the weak scale.
The contributions to these coefficients  from R--parity violation
will be discussed in detail in the following sections.

\section{$CP$ asymmetries with lepton number violation}

\subsection{$\bf \lambda'_{n3j}\lambda'^*_{n2j}:$ Direct $CP$ asymmetry }

With the nonvanishing combination of the R--parity violating couplings 
$\lambda'_{n3j}\lambda'^*_{n2j}$,  the effective Hamiltonian in
Eq.~(\ref{Heff}) includes only the operators of 
the left-handed type: $O^{d_j}_{6L}$ induced by the tree-level 
diagram exchanging sneutrino ${\tilde{\nu}_n}$.   
In addition to these, there arises also additional effective semileptonic
operators,
\begin{equation} \label{O9L}
O^{\nu_n}_{9L}=\overline{s_L}\gamma^\mu b_L \,
\overline{\nu_{nL}}\gamma_\mu \nu_{nL}\,,
\end{equation}
through the tree diagram exchanging the right-handed down squark
${\tilde{d^c_j}}$.
The corresponding Wilson coefficients are given by
\begin{eqnarray} \label{CLRp}
C^{d_j}_{6L}(m_W)&=&
\frac{-1}{4\sqrt{2}G_F \lambda_t}
\sum_{n=1}^3  
\frac{\lambda'_{n3j}\lambda'^*_{n2j}}{m^2_{\tilde{\nu}_n} } \,, \nonumber \\
C^{\nu_n}_{9L}(m_W)&=&
\frac{1}{4\sqrt{2}G_F \lambda_t}
\sum_{j=1}^3  
\frac{\lambda'_{n3j}\lambda'^*_{n2j}}{ m^2_{\tilde{d^c_j}} } \,. 
\end{eqnarray}
Note that $C^{d_j}_{5L}(m_W)=0$ and it remains vanishing 
at low energy scale under one-loop RG evolution as can be seen from 
Eq.~(A9) in the Appendix.
The R--parity violating contributions to the coefficients of the operators 
$O_{7L}$ and $O_{8L}$ come from one-loop diagrams exchanging sneutrinos and
right-handed down squarks, and have been computed in Ref.~\cite{carlos}.
They can be conveniently re-written in terms of the coefficients 
$C^{d_j}_{6L}$ and $C^{\nu_n}_{9L}$ as follows:
\begin{equation} \label{C78Rp}
C_{7L}^{\rpv}(m_W)=  Q_dC_{8L}^{\rpv}(m_W) =
\frac{Q_d}{6}  \left[\sum_n C^{\nu_n}_{9L}(m_W)+
2\sum_j C^{d_j}_{6L}(m_W)\right] \,,
\end{equation}
where $Q_d=-1/3$.  In deriving $C_{7L,8L}^{\rpv}$, we neglected
the down-type squark mixing and the masses of down-type quarks 
compared to the mass of the sneutrino. 
As shown in Eq.~(\ref{yyz}), the effective coefficient $C^{eff}_{7L}$
at $\mu=m_W$ gets a nontrivial contribution from $C^{d_j}_{6L}$ with 
$j=3$ and thus we have
\begin{eqnarray} \label{C78eff}
C^{eff}_{7L}(m_W) &=& C_{7L}(m_W) - C^b_{6L}(m_W)\,,\quad  \nonumber\\
C^{eff}_{8L}(m_W) &=& C_{8L}(m_W) \,.
\end{eqnarray}
{}Making use of the relation (\ref{C78Rp}) and the formula (\ref{A78L})
in the Appendix, one can obtain the Wilson 
coefficients at the scale $\mu =m_b$ as follows;
\begin{eqnarray}  \label{C78mb}
C_{7L}^{eff}(m_b) &=& 0.67\,C^{\rm SSM}_{7L}(m_W)
 + 0.092\,C^{\rm SSM}_{8L}(m_W) -0.17\,C_{2L}(m_W)  \nonumber\\
&-& 0.14[C^d_{6L}(m_W)+C^s_{6L}(m_W)]-0.80 C_{6L}^{b}(m_W) 
 -0.022\, C_{9L}^{\nu_n}(m_W)\,, \nonumber\\
C_{8L}^{eff}(m_b) &=& 0.70\,C^{\rm SSM}_{8L}(m_W)
 -0.080\,C_{2L}(m_W) \nonumber \\
&+&0.42[C_{6L}^{d}(m_W)+C_{6L}^{s}(m_W)]+0.60C_{6L}^{b}(m_W) 
 +0.12\,C_{9L}^{\nu_n}(m_W)  \,,
\end{eqnarray}
where $C^{\rm SSM}_{7L,8L}(m_W)$ contain the contributions from 
the R--parity conserving MSSM sector as well as from the SM one.
We assume that $C_{2L}(m_W)$ comes solely from the SM.
To quantify $C^{\rm SSM}_{7L,8L}(m_W)$ for our purpose,
we introduce the parameters $\eta_{7,8}$ which are defined by
\begin{equation}
C^{\rm SSM}_{7L}(m_W)\equiv \eta_7\,C^{\rm SM}_{7L}(m_W)\,,
\hspace{0.3 cm}
C^{\rm SSM}_{8L}(m_W)\equiv \eta_8\,C^{\rm SM}_{8L}(m_W)\,.
\end{equation}
The parameters $\eta_{7,8}$ are complex in general.

Given the Wilson coefficients in Eq.~(\ref{C78mb}), we are ready to
analyze  the $CP$ violating effects from R--parity violation in 
the  radiative $B$ decay. 
Referring to the work by Kagan and Neubert \cite{kagan} for details,
the direct $CP$ asymmetry $A_{CP}$ and the branching ratios for the
decays $b\to s \gamma$ and $b\to s g$ are given by
\footnote{Here, 
we neglect the R--parity violating contributions to $A_{CP}$ through
terms such as $\Im[C^{\rpv}_{L}C_{7L}^*]$.
}
\begin{eqnarray} \label{OBS}
&&A_{CP}=\frac{1}{\left|C_{7L}\right|^2}
\left\{1.23\,\Im[C_{2L}C_{7L}^*]
-9.52\,\Im[C_{8L}C_{7L}^*]+0.10\,\Im[C_{2L}C_{8L}^*]
\right\} \,\,\, (\%) \,, \nonumber \\
&&{\rm B}(B\rightarrow X_s\gamma)\approx 
2.57\times 10^{-3} K_{\rm NLO}(\delta)
\,\left(\frac{{\rm B}_{\rm semi}}{0.105}\right) \,,
\nonumber \\
&&{\rm B}(B\rightarrow X_{sg})\approx 0.96\,|C_{8L}|^2 
\,{\rm B}_{\rm semi}\,,
\end{eqnarray}
where the coefficients $C$'s without arguments are understood to be the effective
ones evaluated at the scale $m_b$.   The quantity 
$K_{\rm NLO}(\delta)=|C_7|^2+O(\alpha_s,1/m_b^2)$ contains
the corrections to the leading--order result and the specific forms of the
corrections can be found in Ref.~\cite{kagan}.   
We will take $\delta=0.3$ and 
${\rm B}_{\rm semi}=10.5$ \% for the branching ratio of the semileptonic
decay $B\rightarrow X_c\,e\,\overline{\nu}$.
In this work, we take the following values for
the SM predictions for the $C^{\rm SM}_{2L,7L,8L}$ 
at the $m_W$ scale;
\begin{equation}
C_{2L}^{\rm SM}(m_W)\approx  1.0 \,,\hspace{1 cm}
C_{7L}^{\rm SM}(m_W)\approx -0.20 \,,\hspace{1 cm}
C_{8L}^{\rm SM}(m_W)\approx -0.10 \,.
\end{equation}
The above choice of parameters yields the values at the scale 
$\mu=m_b=4.8$ GeV;
\begin{equation}
C_{2L}^{\rm SM}(m_b)\approx  1.11 \,,\hspace{1 cm}
C_{7L}^{\rm SM}(m_b)\approx -0.32 \,,\hspace{1 cm}
C_{8L}^{\rm SM}(m_b)\approx -0.15 \,.
\end{equation}
Considering the above values of the SM coefficients,
Eq.~(\ref{C78mb}) suggests that a significant contribution 
from R--parity violation to the $b\to s \gamma$
decay can arise for  $|C^{\nu_n}_{9L}(m_W)| \sim 10$, 
$|C^{d,s}_{6L}(m_W)| \sim 2$, or $|C^{b}_{6L}(m_W)| \sim 0.3$.
Furthermore, as we will show,
if a sizable $|C^{b}_{6L}|\sim 0.3$ is allowed,  
one can get the direct $CP$ asymmetry of the 
order $|A_{CP}|\sim10 ~\%$ satisfying the observed branching 
ratios of $B\rightarrow X_s\gamma$ and $B\rightarrow X_{sg}$ decay.

\medskip

In order to figure out how large $CP$ asymmetry can come from R--parity
violation, let us consider the experimental bounds on the new
Wilson coefficients appeared in Eq.~(\ref{C78mb}).
First of all, those coefficients will be constrained by the experimental data 
for the branching ratios in Eq.~(\ref{OBS}).
In this work, we use the CLEO data:
\begin{eqnarray}
{\rm B}(B\rightarrow X_s\gamma) &=&
(3.15\pm 0.35_{\rm stat}\pm 0.32_{\rm stat} 
\pm 0.26_{\rm model})\times 10^{-4}
\,, 
~~~~\cite{cleo.bsp}\nonumber \\
{\rm B}(B\rightarrow X_{sg}) &\lesssim& 6.8 ~\% ~~{\rm (90 \% C.L.)}\,.
~~~~\cite{cleo.bsg} \nonumber
\end{eqnarray}
More important constraints on the coefficients $C^{d_j}_{6L}$ and 
$C^{\nu_n}_{9L}$ come from experimental data on the various B meson decays.
Let us discuss the relevant bound for each coefficient.

First, the coefficient $C^{d}_{6L}$ is constrained by
the $B$ decay mode $\overline{B^0}\rightarrow K^0\pi^0$,
whose branching ratio is observed to be \cite{PDG}
$${\rm B}(\overline{B^0}\rightarrow K^0\pi^0) < 4.1 \times 10^{-5}\,.
$$
Following the similar method used in Ref.~\cite{aliko}, 
we estimate the matrix element of the operator $O^{d}_{6L}$ as
\begin{equation}
<K^0\pi^0|O^{d}_{6L}|\overline{B^0}>\approx
i\frac{ m_K^2(m_B^2-m_\pi^2)}{2(m_s+m_d)(m_b-m_d)} f_K
F_1^{B\rightarrow\pi}(m_K^2) \,.
\end{equation}
Taking $f_K=0.16$ GeV, $F_1^{B\rightarrow\pi}(m_K^2)=0.33$, we obtain
\begin{eqnarray}
\left|C^{d}_{6L}\right| < 0.17 \,.
\label{eq:con1}
\end{eqnarray}

Second, the coefficient $C^{s}_{6L}$ is constrained by
considering the $B$ decay mode $\overline{B^0}\rightarrow \phi K^0$.  
The experimental limits on
the branching ratio of this decay mode is \cite{cleo.bkpi}
$${\rm B}(\overline{B^0}\rightarrow \phi K^0) < 3.1 \times 10^{-5}\,.
$$
{}From our estimation of the matrix element of the operator $O^{s}_{6L}$;
\begin{equation}
<\phi K^0|O^{s}_{6L}|\overline{B^0}>\approx
\frac{1}{4N}f_\phi m_\phi 
\,\epsilon\cdot(P_B+P_K)\,F_1^{B\rightarrow K}(m_\phi^2)
\,,
\end{equation}
where $\epsilon$ is a polarization vector of $\phi$ and we will take $N=3$.
With $f_\phi=0.23$ GeV, $F_1^{B\rightarrow K^0}(m_\phi^2)=0.38$,  
we obtain following limit :
\begin{eqnarray}
\left|C^{s}_{6L}\right| < 0.23 \,. 
\label{eq:con2}
\end{eqnarray}
The above bounds in Eqs.~(\ref{eq:con1}) and (\ref{eq:con2}) tell us that
the coefficients $C^{d,s}_{6L}$ cannot play any important
role in the radiative
B decays as can be seen from Eq.~(\ref{C78mb})
\footnote{The bounds in Eqs.~(\ref{eq:con1}) and (\ref{eq:con2}) are at the
scale $m_b$. Practically, the Wilson
coefficients induced by the R--parity violating couplings at the scale $m_b$ are
nearly the same as those at the scale $m_W$ [see Eq.~(A9)].}.

Now let us consider the constraint on $C^{b}_{6L}$.
The most useful bound comes {\it indirectly} from the consideration of the 
decay mode $\overline{B^0}\rightarrow X_s \nu_n \overline{\nu_n}$
whose branching ratio can be calculated as \cite{nardi}
\begin{equation}
{\rm B}(\overline{B^0}\rightarrow X_s \nu_n \overline{\nu_n})
=\left|\frac{\lambda_t}{V_{cb}}\right|^2
\frac{\left|C^{\nu_n}_{9L}\right|^2}{f_{\rm PS}(m_c^2/m_b^2)}\,
{\rm B}_{\rm semi}\,.
\end{equation}
According to the analyses in Ref.~\cite{nardi}, one obtains 
{\it indirect} experimental information on the above branching ratio:
$${\rm B}(\overline{B^0}\rightarrow X_s \nu_n \overline{\nu_n})
<3.9\times 10^{-4}\,.$$  
Taking this value and 
$\left|\lambda_t/V_{cb}\right|=0.976$, $f_{\rm PS}(m_c^2/m_b^2)=0.5$,
we put the bound, 
\begin{equation}  
\label{bsnn}
\left|C^{\nu_n}_{9L}\right| < 0.044 \,.  
\end{equation} 
Thus the contribution of the coefficient $C^{\nu_n}_{9L}$
to Eq.~(\ref{C78mb}) can also be neglected.  
Now, under the condition that the bound (\ref{bsnn}) is applied to
each component of the coefficient [see Eq.~(\ref{CLRp})], we get
\begin{equation}
|C^{b}_{6L}| < 0.044
\,\left(\frac{m^2_{\tilde{d^c_3}}}{m^2_{\tilde{\nu_n}}}\right)\,,
\label{eq:con4}
\end{equation}
for each $n=1,2,3$.  Here we remark that under the condition 
(\ref{bsnn}), the R--parity violating contribution to the
semileptonic decay $B\rightarrow X_c l_n \overline{\nu_n}$ through
the effective operator 
\begin{equation} \label{Hsemi}
{\cal H}_{eff} =  
-{\lambda'_{n3j}\lambda'^*_{n2j} \over 2 m^2_{\tilde{d}^c_j} } 
\overline{c_L}\gamma^\mu  b_L \,
\overline{e}_{nL}\gamma_\mu \nu_{nL}\,,
\end{equation}
can be made small enough to  satisfy the {\it direct} experimental
bounds \cite{Semi};
\begin{eqnarray}     \label{Semi}
|\lambda'_{233}\lambda'^*_{223}|<  1.1\times 10^{-3} 
\left(\frac{m_{\tilde{d^c_3}}}{100 ~{\rm GeV}}\right)^2 \,,
\nonumber \\
|\lambda'_{333}\lambda'^*_{323}|<  4.4\times 10^{-3} 
\left(\frac{m_{\tilde{d^c_3}}}{100 ~{\rm GeV}}\right)^2  \,.
\end{eqnarray}

Finally, we have to consider 
the neutrino mass coming from our choice of nonvanishing 
$\lambda'_{n33}\lambda'^*_{n23}$.
With nonzero value of  $\lambda'_{n33}$, the neutrino
$\nu_n$ may get an undesirably large mass from one-loop diagrams
with the exchange of bottom quark and squark \cite{numass}. 
The one-loop contribution to the neutrino mass is given by
$$
 m_{\nu_n} \approx {3\over 8\pi^2} { \lambda'^2_{n33} 
 A m_b^2 \over \tilde{m}^2 }
$$
where $A$ denotes left-right sbottom mixing parameter and $\tilde{m}$ 
is an average sbottom mass.  Taking $\lambda'^2_{n33}/\tilde{m}^2 < 
10^{-7}{\rm GeV}^{-2}$ from the consideration of Eq.~(\ref{bsnn}), 
we obtain 
\begin{equation}
m_{\nu_n} \lesssim  10\, {\rm keV} 
\end{equation}
with $A=100$ GeV.
Thus it is well below the direct experimental limit for the 
muon (tau) neutrino which is 0.17 (18) MeV \cite{PDG}.
Yet another indirect limit for the muon or tau neutrino mass
comes from the observation of atmospheric muon neutrinos.
The atmospheric neutrino data from the Super-Kamiokande 
are known to be nicely explained by the neutrino oscillation 
between largely mixed muon and tau neutrinos \cite{skam}.
A natural consequence of this would be that the muon and tau neutrinos 
are very light $m_{\nu_\mu,\nu_\tau} \lesssim 1$ eV.
If this is the case, there is no room at all for large $CP$ asymmetry 
from R--parity violation.
However, having a large R--parity violation and thus
a heavier muon or tau neutrino is not excluded completely
as there exist some other viable options for the 
explanation of the Super-Kamiokande data, such as
a neutrino decay \cite{ndecay}.

{}From the above consideration, 
the only possibility for a significant enhancement of 
R--parity violating contribution to the $b\to s \gamma$ decay
is to have a large coefficient $C_{6L}^{b}$ with a 
sfermion mass hierarchy $m_{\tilde{d}^c_3} > m_{\tilde{\nu}_n}$.
For example, we need  $m_{\tilde{d}^c_3}\sim 5\,m_{\tilde{\nu}_n}$
to get $C_{6L}^{b}\sim 1$. 
Taking into account all the above experimental limits on the 
Wilson coefficients, let us now analyze how large $CP$ 
asymmetry $A_{CP}$ can be obtained.
In FIG.~2, we show $A_{CP}$ as a function of the branching ratio of 
the decay $B \rightarrow X_s\gamma$ varying $|C_{6L}^{b}|$ and 
${\rm Arg}(C_{6L}^{b})$ from 0 to $2\pi$. The other R--parity violating
couplings are neglected.
We take $\eta_7=\eta_8=\eta$ as a real number in FIG.~2 even though 
$\eta$ is a complex number generally.  In this figure we consider
additional
experimental constraints coming from B($B \rightarrow X_s\gamma$),
B($B \rightarrow X_{sg}$). 
If $m^2_{\tilde{d}^c_3}$ are larger than $m^2_{\tilde{\nu}_n}$, then
sizable $|C_{6L}^{b}|$ is allowed evading the bounds Eq.~(\ref{eq:con4}).
We find the $CP$ asymmetry can reach $13~\%$ for 
$m_{\tilde{d}^c_3}/m_{\tilde{\nu}_n} \approx 3.4$ with vanishing
R--parity conserving supersymmetric contributions ($\eta=1$)
as shown in the left-upper frame of FIG.~2. 
In the case where the R--parity conserving
supersymmetric contributions take the same sign as the SM values
of $C_{7L,8L}(m_W)$, this $CP$ asymmetry can be larger as shown
in FIG.~2 with $\eta=2$.
On the other hand, the $CP$ asymmetry decreases
when the R--parity conserving
supersymmetric contributions take the opposite sign to the SM values
of $C_{7L,8L}(m_W)$ as seen from FIG.~2 with $\eta=0,-1$.

\medskip

Before concluding this subsection, it is worthwhile to notice
that the $CP$ asymmetry in the hadronic $B$ decays 
such as $B_d\to \pi K_S$ and $B_d\to\phi K_S$ can be significantly affected by
the R--parity violating couplings $C_{6L}^{d,s}$ even if the effects of
the couplings on the direct $CP$ asymmetry in the radiative $B$ decay are negeligible
\cite{guetta,jang,datta}.


\subsection{$\bf \lambda'_{nj2}\lambda'^*_{nj3}:$
     Mixing-induced $CP$ asymmetry}

Contrary to the previous case, the combination of R--parity violating
couplings $\lambda'_{nj2}\lambda'^*_{nj3}$ leads only to the 
right-handed set of operators: $O_{7R,8R}$ and $O^q_{5R,6R}$
in the effective Hamiltonian (\ref{Heff}).  In addition, the coefficient
of the operator $O_{6R}^t$ is also generated. Even though it
does not contribute to the $b$ decay,  it's coefficient will be considered
since it enters into $C_{7R,8R}$.  As in the previous subsection,
we have also the semileptonic four-Fermi operators as follows:
\begin{eqnarray} \label{qqllR}
O^{\nu_n}_{9R}&=&\overline{s^\alpha_R}\gamma^\mu b^\alpha_R \,
\overline{\nu_{nL}}\gamma_\mu\nu_{nL}\,, \nonumber\\
O^{e_n}_{10R}&=&\overline{s^\alpha_R}\gamma^\mu b^\alpha_R \,
\overline{e_{nL}}\gamma_\mu e_{nL}\,.
\end{eqnarray}
The nonzero Wilson coefficients at $m_W$ induced
from our R--parity violation are
\begin{eqnarray} \label{CRRp}
C^{q_j}_{6R}(m_W)&=& \frac{-1}{4\sqrt{2}G_F \lambda_t}
\sum_{n=1}^3\frac{\lambda'_{nj2}\lambda'^*_{nj3}}
{m^2_{\tilde{l}_n}} \,, \nonumber \\
C^{\nu_n}_{9R}(m_W)&=& C^{e_n}_{10R}(m_W) =
\frac{1}{4\sqrt{2}G_F \lambda_t}
\sum_{j=1}^3\frac{\lambda'_{nj2}\lambda'^*_{nj3}}{m^2_{\tilde{Q}_j}} \,, 
\end{eqnarray}
where $q_j$ can be either $u_j$ or $d_j$, $m_{\tilde{l}_n}$ and 
$m_{\tilde{Q}_j}$ are the masses of the doublet slepton and squark, 
respectively.
In deriving these coefficients, 
we do not consider the effect of the CKM mixing.
Note that there can be also other four-Fermi operators
involving two different flavors of quarks or leptons through
the CKM mixing, which we neglect in our discussion as they give
subleading contributions.
Now, disregarding the contributions from the R--parity conserving 
supersymmetric sector, the Wilson coefficients for $O_{7R,8R}$ 
at the scale $m_W$ \cite{carlos} can be expressed in
terms of the coefficients in Eq.~(\ref{CRRp}) as follows 
\begin{eqnarray} \label{C7RRp}
C_{7R}(m_W)=C_{7R}^{\rpv}(m_W) &= &  
\frac{Q_d}{6} \left[\sum_n C^{\nu_n}_{9R}(m_W)
+2\sum_j C^{q_j}_{6R}(m_W) \right] \nonumber\\
&& + {1\over 18} \left[ 8 \sum_n C^{e_n}_{10R}(m_W)
+7\sum_j C^{q_j}_{6R}(m_W) \,P_\gamma(x_j) \right]  \,, \\
C_{8R}(m_W)=C_{8R}^{\rpv}(m_W) &= &  
{1\over6} \left[\sum_n C^{\nu_n}_{9R}(m_W)
+2\sum_j C^{q_j}_{6R}(m_W) \right. \nonumber\\
&& +  \left.\sum_n C^{e_n}_{10R}(m_W)
+2\sum_j C^{q_j}_{6R}(m_W) \,P_g(x_j) \right]  \,,
\label{C8RRp}
\end{eqnarray}
where 
$$P_{\gamma}(x_j)={36 \over 7} [Q_u F_1(x_j)+F_2(x_j)]\,,\quad
        P_g(x_j) = 6 F_1(x_j)$$
with $x_j=m^2_{u_j}/m^2_{\tilde{l}_n}$.
The functions $F_{1,2}$ are defined as
\begin{eqnarray}
 F_1(x)&=&{1\over12} {(2+3x-6x^2+x^3+6x\ln x)\over(1-x)^4} \,,\nonumber\\
 F_2(x)&=&{1\over12} {(1-6x+3x^2+2x^3-6x^2\ln x)\over(1-x)^4} \,.\nonumber
\end{eqnarray}
Note that these functions are defined as $P_\gamma(0)=P_g(0)=1$
and their values for some selected $x$ are listed in TABLE~II

\medskip

In the case under consideration, 
there is no phase in $C_{7R}C_{8R}^*$ contributing
the direct $CP$ asymmetry in right-handed sector analogue of 
Eq.~(\ref{OBS})
since the same combination of the couplings generates both $C_{7R,8R}$.
However, there can arise a sizable mixing-induced $CP$ asymmetry as 
defined in Eq.~(\ref{AM}).  
That is, we may have $|A_M| \approx 1$ with  $|C_{7L}| \approx |C_{7R}|$.  
Here,  the Wilson coefficients are 
effective ones evaluated at the scale $m_b$.  
Combining our results in the Appendix (\ref{A78R}) 
and Eq.~(\ref{C7RRp}), we find 
\begin{eqnarray} \label{C7Rmb}
C^{eff}_{7R}(m_b) &=&  0.40\,[C^{q_1}_{6R}(m_W)+C^{q_2}_{6R}(m_W)]       
+[-0.80+0.26\,P_\gamma(x_t) +0.031\,P_g(x_t)]\,C^{q_3}_{6R}(m_W) 
\nonumber \\
&& -0.022\,C^{\nu_n}_{9R}(m_W) +0.31\,C^{e_n}_{10R}(m_W) \,.
%
\end{eqnarray}
{}From this equation, it appears that we can easily obtain 
$C_{7R} \approx  |C_{7L}|=|C^{\rm SSM}_{7L}|\approx 0.32\,|\eta_7|$.
To clarify this, we now consider the experimental constraints
on the coefficients appearing in Eq.~(\ref{C7Rmb}). 
Let us first note that the arguments in the previous subsection are
applied here to get the bounds; 
\begin{equation}
|C^{q_1}_{6R}|<0.17\,,\quad
|C^{q_2}_{6R}|<0.23 \,,\quad
|C^{\nu_n}_{9R}|<0.044
\end{equation}
as in Eqs.~(\ref{eq:con1}), (\ref{eq:con2}) and (\ref{bsnn}), respectively.  
Thus, the contributions of these coefficients
in Eq.~(\ref{C7Rmb}) are not significant.
Another important constraint on the relevant R--parity violating
couplings comes from the decay $B\rightarrow X_s l^+_n l^-_n$ 
\cite{jkl} induced by the second effective operator in Eq.~(\ref{qqllR}).
This consideration leads to 
\begin{equation}
\label{conR}
|C_{10R}^{e_n}|<0.017\,
\left(\sqrt{\frac{{\rm B}(b\rightarrow X_s l^+_n l^-_n)^{\rm expt}}
{5.7\times 10^{-5}}}\right)\,,
\end{equation}
where ${\rm B}(b\rightarrow X_s l^+_n l^-_n)^{\rm expt}$ denotes the
experimental upper limit on the branching ratio of the 
$B\rightarrow X_s l^+_n l^-_n$ decay modes given by 
\cite{PDG}
\begin{eqnarray}
&&{\rm B}(B\rightarrow X_s e^+ e^-) < 5.7\times 10^{-5} \,,
\nonumber \\ \nonumber
&&{\rm B}(B\rightarrow X_s \mu^+ \mu^-) < 5.8\times 10^{-5} \,.
\end{eqnarray}
Even if there are no data for $b\rightarrow X_s \tau^+ \tau^-$, 
we expect it's branching ratio is most probably less than that of 
$B\rightarrow X_s e^+ e^-$.  
Then, similarly to Eq.~(\ref{eq:con4}),
the bounds on $C^q_{6R}$ can be obtained 
indirectly from Eq.~(\ref{conR}) as
\begin{equation}
|C_{6R}^{q_j}|<0.017\,
\left(\sqrt{\frac{{\rm B}(B\rightarrow X_s l^+_n l^-_n)^{\rm expt}}
{5.7\times 10^{-5}}}\right)
\left(\frac{m_{\tilde{Q}_j}}{m_{\tilde{l}_n}}\right)^2\,,
\end{equation}
for each $j$ and $n$.
Therefore, we may obtain a sizable $C^{eff}_{7R}(m_b)$ from the 
contribution of $C^{q_3}_{6R}(m_W)$ if there is again a hierarchy between 
sfermion masses.  For example, taking $m_{\tilde{Q}_3}/m_{\tilde{l}_n}
\approx 5$, $|C_{6R}^{q_3}|=0.33$ 
is allowed within the present experimental bound.
Taking $x_t=1$ which gives $P_{\gamma}(1)=5/14$ and $P_g(1)=1/4$,
one could obtain $|C_{7R}^{eff}(m_b)|\approx 0.23$. 
If $|C_{7L}^{eff}(m_b)|=0.23$, $|A_M|\approx 1$ is possible accommodating 
the measured ${\rm B}(b\rightarrow s\gamma)$.
Finally, let us note that a large mixing-induced $CP$ asymmetry 
requires the (tau) neutrino to be heavy as discussed 
in the previous subsection.

\medskip

Again we note that with the coupling $\lambda'_{n22}\lambda'^*_{n32}$
which does not affect the radiative $B$ decays, 
one can have important 
effects on the $CP$ asymmetries in the hadronic $B$ decays such as
$B_d \to \phi K_S$ \cite{guetta,jang} 
and $B^\pm \to \pi^\pm K^0$ \cite{datta}.

\section{$CP$ asymmetries with baryon number violation}

\subsection*{$\bf \lambda''_{n12}\lambda''^*_{n13}:$ 
             Mixing-induced $CP$ asymmetry}

Our final case is to have baryon number violation while
lepton number is conserved.   Then, the new operator set for the
$b\to s $ transition contains  again only right-handed ones
with nonvanishing product of couplings
$\lambda''_{n12}\lambda''^*_{n13}$: $O^{u_n,d}_{3R,4R}$.  
The Wilson coefficients of these operators 
calculated at the weak scale $m_W$ are
\begin{eqnarray} \label{CBRp}
C^{u_n}_{3R}(m_W)&=& \frac{1}{4\sqrt{2}G_F \lambda_t}
\frac{\lambda''_{n12}\lambda''^*_{n13}  }
{m^2_{\tilde{d}^c_1}} =-C_{4R}^{u_n}\,, \nonumber \\
C^{d}_{3R}(m_W)&=& \frac{1}{4\sqrt{2}G_F \lambda_t}
\sum_{n=1}^3\frac{\lambda''_{n12}\lambda''^*_{n13}}
{m^2_{\tilde{u}^c_n}}= -C_{4R}^{d} \,.
\end{eqnarray}
Notice that the simultaneous presence of nonvanishing coefficients
$C_{3R,4R}$ is due to color antisymmetry in the superpotential
term, $U^cD^cD^c$.  Following the similar steps as before, we get
the relation 
\begin{eqnarray} \label{C78BRp}
C_{7R}(m_W)=
C_{7R}^{\rpv}(m_W)&=&  {1\over9} \left[ 4C^{d}_{4R}(m_W) 
        -5\sum_n C^{u_n}_{4R}(m_W) P'_\gamma(x_n) \right]  \,, 
  \nonumber\\
C_{8R}(m_W)=
C_{8R}^{\rpv}(m_W)&=&  {1\over6} \left[ C^{d}_{4R}(m_W) + 
          \sum_n C^{u_n}_{4R}(m_W) P'_g(x_n) \right]  \,, 
\end{eqnarray}
where 
$$ P'_\gamma(x_n)={36\over 5}[Q_u F_1(x_n)-Q_d F_2(x_n)]\,,\quad
   P'_g(x_n)=12[F_1(x_n)-F_2(x_n)] $$
with $x_n=m_{u_n}^2/m^2_{\tilde{d}^c}$ and $P'_\gamma(0)=P'_g(0)=1$.
The Wilson coefficient $C^{eff}_{7R}$ for the $b\to s\gamma$ decay 
at $m_b$ is
\begin{eqnarray} \label{C7Bmb}
C^{eff}_{7R}(m_b)&=& 0.21\, C^d_{4R}(m_W)
-0.17\, [C^{u}_{4R}(m_W)+ C^{c}_{4R}(m_W)] \nonumber\\
&&-[0.37\,P'_{\gamma}(x_t)-0.015\,P'_g(x_t)]\,C^t_{4R}(m_W) \,.
\end{eqnarray}

Let us now consider the experimental limits for the various
coefficients in Eq.~(\ref{C7Bmb}).
First of all, the R--parity violating couplings with $n=1$
are strongly constrained by the non-observation of nucleon-antinucleon
oscillation and double nucleon decay; $\lambda''_{113} <5\times10^{-3}$ 
and $\lambda''_{112} <10^{-6}$ with sfermion mass of 300 GeV \cite{NN}. 
Thus $C^u_{4R}$ are negeligibly small.
The constraint for $C^{d}_{4R}(m_W)$ comes again from 
the $B\to K^0\pi^0$ decay. We estimate the matrix element of the operator
$O_{4R}^d$ as
\begin{equation}
<K^0\pi^0|O^{d}_{4R}|\overline{B^0}>\approx
i\,(m_B^2-m_\pi^2)\, f_K\,
F_1^{B\rightarrow\pi}(m_K^2) \,.
\end{equation}
Using the similar values used in Eq.~(\ref{eq:con1}), we obtain
\begin{equation} \label{CdR}
|C^{d}_{4R}|< 0.15\,.
\end{equation}
Note that this bound is also consistent with the data for the
decay mode $B^+\to \bar{K}^0\pi^+$ \cite{carl}.
Concerning the coefficient $C^{c}_{4R}$, the consideration 
of the $B\rightarrow J/\psi K_S$ decay gives \cite{jang} 
\begin{equation}
|C^{c}_{4R}|< 0.02\,.
\end{equation}
Applying again the bound (\ref{CdR}) to each component of $C^d_{4R}$,
we get the bound on the $C^t_{4R}$;
\begin{equation}
|C^{t}_{4R}|< 0.15\, { m^2_{\tilde{t}^c} \over m^2_{\tilde{d}^c} } \,.
\end{equation}
%
%
Thus, to get $|C^{eff}_{7R}|>0.2$  leading to a nearly maximal
mixing-induced $CP$ asymmetry, we need a sizable $C^t_{4R}$
which can come about for $m_{\tilde{t}^c} >3.5 \,m_{\tilde{d}^c}$ 
with $x_t=1$ 
\footnote{The constraint on the relevant single baryon number violating
coupling comes from $\Gamma(Z\to l\bar{l})/\Gamma(Z\to {\rm hadrons})$ 
\cite{ZZ}, which gives $|\lambda''_{312,313,323}| <0.5 \,$
for $\tilde{m}=100$ GeV. This constraint is so weak that
$|C^t_{3R,4R}|\sim{\cal O}(1)$ is easily allowed.}.

\medskip

We conclude this section by pointing out that the  direct $CP$ asymmetry 
in the decay $B^\pm \to \pi^\pm K^0$ can arise  maximally as is the 
case with the coupling $\lambda'_{n22}\lambda'^*_{n32}$ \cite{datta}.

\section{Conclusion}

We have discussed the effects of R--parity violation on the 
$CP$ asymmetries in radiative $B$ decays. 
When we allow R--parity and lepton number violating couplings which
generate at one-loop level the tau neutrino mass of order 10 keV,
they can induce rather large $CP$ violating effects in the 
$b\to s \gamma$ decay.  The direct $CP$ asymmetry can be
as large as 17 \% if the R--parity conserving supersymmetric contribution is
comparable to the SM one.  The  mixing-induced $CP$ asymmetry
can be almost maximal depending on the sfermion masses.
For these sizable $CP$ violating effects, it is required to have
moderate sfermion mass splittings by factor 4 or bigger.
If the atmospheric neutrino data from the Super-Kamiokande 
are to be explained by the oscillation between the muon and tau neutrinos 
whose masses are very light $m_{\nu_\mu,\nu_\tau} \lesssim 1$ eV,
then the effects of R--parity violation on the radiative $B$ decays are
negligible.  
A large mixing-induced $CP$ asymmetry is also possible with the R--parity 
and baryon number violating couplings with the similar order of 
sfermion mass splitting.

These results could be contrasted with the R--parity violating
effects on the $CP$ asymmetries in the hadronic $B$ decays 
such as $B_d\to J/\psi K, \phi K$ or $B^\pm \to \pi^\pm K$,
which could be significant without sfermion mass splitting and are 
rather insensitive to the neutrino mass restriction.

\bigskip

{\it Note added}:  While our work was being prepared, we encountered 
the paper \cite{zurich} which considers the R--parity violating effect
on the radiative $B$ decay.  It also deals with the RG running of 
the enlarged set of the Wilson coefficients which overlaps partly with 
our paper, and we find discrepancies in anomalous dimension matrix 
elements and the relation like (A10) and (A11).

\section*{Acknowledgments}
We thank G. Bhattacharyya for helpful discussions.
The work of  KH was supported by grant No.~1999-2-111-002-5 from
the interdisciplinary Research program of the KOSEF and BK21 project of the
Ministry of Education.

\appendix
\section{RG Equations with the extended operators}
\def\f#1#2{\frac{#1}{#2}}
Here, we present the anomalous dimension matrix for the 28 operators
including the standard set $O_{1,\cdots8}$ and the extended set 
$O^q_{3,\cdots,6}$ with $q=u,d,s,c,b$.  We drop the indices $L,R$ for
the left-handed and right-handed set of operators as they have
the identical anomalous dimension matrix.  Omitting the usual
$8\times8$ matrix for the standard set of operators, we have the following
nonvanishing block-diagonal elements of the whole $28\times28$ matrix.

The $2\times2$ submatrices mixing the operators $O^q_{3,4}$ and 
$O^q_{5,6}$ with themselves are
\begin{equation} \label{ADM1}
\pmatrix{ -2 & 6 \cr 6 & -2 \cr}\quad{\rm and}\quad
\pmatrix{ 2 & -6 \cr 0 & -16 \cr}\,,
\end{equation}
respectively.  
The $2\times 8$ submatrix mixing the operators
$O^q_{3,4}$ with the standard set $O_{1,\cdots,8}$ is given by
\begin{equation} \label{ADM2}
\pmatrix{ 0 & 0 & -{2\over9}\delta_{qb,s} & {2\over3} \delta_{qb,s} &
          -{2\over9}\delta_{qb,s} & {2\over3} \delta_{qb,s} &
          -{232\over81}\delta_{qb,s} & 3+ {70\over27}\delta_{qb,s} \cr
         0 & 0 & -{2\over9} & {2\over3}  & -{2\over9} & {2\over3}  &
         {416\over81}\delta_{qu}-{232\over81}\delta_{qd}  & 
         {70\over27} + 3\delta_{qb,s} \cr }
\end{equation}
where 
\begin{eqnarray}
&&\delta_{qb,s}= 
\cases{1 \quad\mbox{for $q=b$ or $s$} \cr 
       0 \quad\mbox{otherwise,} \cr  }  \nonumber\\
&&\delta_{qu}, \delta_{qd}= 
\cases{1 \quad\mbox{when $q$ is an up-type, or down-type quark}  \cr 
       0 \quad\mbox{otherwise.} \cr } \nonumber
\end{eqnarray}
Similarly, the $2\times 8$ submatrix mixing the operators
$O^q_{5,6}$ with the 8 standard operators is 
\begin{equation} \label{ADM3}
\pmatrix{ 0 & 0 & 0 & 0 & 0 & 0 &
          {32\over9}\delta_{qb} & -3 -{14\over3}\delta_{qb}  \cr
         0 & 0 & -{2\over9} & {2\over3}  & -{2\over9} & {2\over3}  &
         -{448\over81}\delta_{qu}+{200\over81}\delta_{qd}  & 
         -{119\over27} - 4\delta_{qb} \cr }
\end{equation}
where 
$$\delta_{qb}= 
\cases{1 \quad\mbox{for $q=b$ } \cr 
       0 \quad\mbox{otherwise.} \cr  } $$

With this anomalous dimension matrix, the low energy Wilson coefficients
are given by
\begin{equation}\label{rgcu}
\vec C^{eff}(\mu)=\hat U^{eff}(\mu, \mu_W) \vec C^{eff}(\mu_W)
\end{equation}
where
\begin{equation}\label{u0vd0}
\hat U^{eff}(\mu,\mu_W)= \hat V
\left({\left[{\alpha_s(\mu_W)\over\alpha_s(\mu)}
\right]}^{{\vec\gamma^{eff}\over 2 \beta_0}}
   \right)_D \hat V^{-1}   \end{equation}
where $\hat V$ diagonalizes ${\hat\gamma^{eff T}}$
\begin{equation}\label{ga0d}
\hat\gamma^{eff}_D=\hat V^{-1} {\hat\gamma^{eff T}} \hat V
  \end{equation}
and $\vec\gamma^{eff}$ is the vector containing the diagonal elements of
the diagonal matrix of eigenvalues of $\hat{\gamma}^{eff}$.

By solving Eqs.~(A4)--(A6), we find the following 
low energy Wilson coefficients at $\mu_b$ in terms of 
nonzero $C_{6}^{d_i}(\mu_W),C_{6}^{u,c}(\mu_W)$   in the 
L or R sector induced at $\mu_W$ by $R$ parity violating coupling 
$\lambda'_{n2j}\lambda'^*_{n3j}$ or  $\lambda'_{nj2}\lambda'^*_{nj3}$,
respectively, and $C_{3,4}^{u,d,c}(\mu_W)$ in the R sector by  
$\lambda''_{n12}\lambda''^*_{n13}$,
and the SM contribution $C_{2L}(\mu_W)$ in L sector:
\begin{eqnarray}
C_{7}^{eff}(\mu_b) &=& \eta^{16/23} C_{7}^{eff}(\mu_W)
 + {8\over 3}\left(\eta^{14/23}-\eta^{16/23}\right) C_{8}^{eff}(\mu_W)
 + C_2(\mu_W) \sum^{10}_{i=1}h_i\eta^{a_i}  \nonumber \\
&+&\left(C_{6}^{d}(\mu_W)+C_{6}^{s}(\mu_W)\right) 
    \sum^{10}_{i=1}r^1_i\eta^{a_i} 
 + C_{6}^{b}(\mu_W) \sum^{10}_{i=1}r^2_i\eta^{a_i} 
 + \left(C_{6}^{u}(\mu_W)+C_{6}^{c}(\mu_W)\right) \sum^{10}_{i=1}r^3_i\eta^{a_i} \nonumber \\ 
&+&C_{3}^{d}(\mu_W) \sum^{10}_{i=1}r^4_i\eta^{a_i} 
 + C_{4}^{d}(\mu_W) \sum^{10}_{i=1}r^5_i\eta^{a_i}  \nonumber \\
&+&\left( C_{3}^{u}(\mu_W) + C_{3}^{c}(\mu_W) \right) 
   \sum^{10}_{i=1}r^6_i\eta^{a_i} 
 + \left( C_{4}^{u}(\mu_W) + C_{4}^{c}(\mu_W) \right) 
    \sum^{10}_{i=1}r^7_i\eta^{a_i}  \\
C_{8}^{eff}(\mu_b) &=& \eta^{14/23} C_{8}^{eff}(\mu_W)
 + C_2(\mu_W) \sum^{10}_{i=1}\bar{h}_i\eta^{a_i}  \nonumber \\
&+&\left(C_{6}^{d}(\mu_W)+C_{6}^{s}(\mu_W)\right) 
    \sum^{10}_{i=1}\bar{r}^1_i\eta^{a_i} 
 + C_{6}^{b}(\mu_W) \sum^{10}_{i=1}\bar{r}^2_i\eta^{a_i} 
 + \left(C_{6}^{u}(\mu_W)+C_{6}^{c}(\mu_W)\right) 
   \sum^{10}_{i=1}\bar{r}^3_i\eta^{a_i} \nonumber \\ 
&+&C_{3}^{d}(\mu_W) \sum^{10}_{i=1}\bar{r}^4_i\eta^{a_i} 
 + C_{4}^{d}(\mu_W) \sum^{10}_{i=1}\bar{r}^5_i\eta^{a_i}  \nonumber \\
&+&\left( C_{3}^{u}(\mu_W) + C_{3}^{c}(\mu_W) \right) 
    \sum^{10}_{i=1}\bar{r}^6_i\eta^{a_i} 
 + \left( C_{4}^{u}(\mu_W) + C_{4}^{c}(\mu_W) \right) 
    \sum^{10}_{i=1}\bar{r}^7_i\eta^{a_i}  \\
C_2(\mu_b) &=& {1\over2}\left(\eta^{6/23}+\eta^{-12/23}\right) C_2(\mu_W) 
    \nonumber \\
C_{5}^{d_i,u,c}(\mu_b) &=& 0 \nonumber \\
C_{6}^{d_i,u,c}(\mu_b) &=&  C_{6}^{d_i,u,c}(\mu_W) \nonumber\\
C_3^{u,d,c}(\mu_b) &=& {1\over2}\left(\eta^{6/23}+\eta^{-12/23}\right) 
    C_3^{u,d,c}(\mu_W)
+ {1\over2}\left(\eta^{6/23}-\eta^{-12/23}\right) 
   C_4^{u,d,c}(\mu_W) \nonumber \\
C_4^{u,d,c}(\mu_b) &=& {1\over2}\left(\eta^{6/23}+\eta^{-12/23}\right) 
    C_4^{u,d,c}(\mu_W)
+ {1\over2}\left(\eta^{6/23}-\eta^{-12/23}\right) C_3^{u,d,c}(\mu_W) 
\end{eqnarray}
where the quantities $a_i, h_i,\bar{h}_i, r^n_i, \bar{r}^n_i$ are shown in the
TABLE~I.

The explicit numerical expressions with the choice $\mu_W=m_W$, 
$\mu_b=4.8$ GeV, $\alpha_s(\mu_W)=0.120$ and $\alpha_s(\mu_b)=0.214$
are 
\begin{eqnarray} 
\label{A78L}
C_{7L}^{eff}(\mu_b) &=& 0.6687 C_{7L}(\mu_W)
 + 0.0920 C_{8L}(\mu_W)
 -0.1732 C_{2L}(\mu_W)  \nonumber\\
&-&0.0974  \left( C_{6L}^{d}(\mu_W) + C_{6L}^{s}(\mu_W)\right) 
 -(0.6687+0.0875) C_{6L}^{b}(\mu_W) \nonumber \\
C_{8L}^{eff}(\mu_b) &=& 0.7032 C_{8L}(\mu_W)
 -0.0801 C_{2L}(\mu_W) \nonumber \\
&+&0.1893  \left( C_{6L}^{d}(\mu_W)+C_{6L}^{s}(\mu_W)\right) 
 +0.3670 C_{6L}^{b}(\mu_W) \\
\label{A78R}
C_{7R}^{eff}(\mu_b) &=& 0.6687 C_{7R}(\mu_W)
 + 0.0920 C_{8R}(\mu_W) \nonumber\\
&-&0.0974  \left( C_{6R}^{d}(\mu_W) + C_{6R}^{s}(\mu_W)\right) 
 -(0.6687+0.0875) C_{6R}^{b}(\mu_W) \nonumber \\
&+&0.2506 \left( C_{6R}^{u}(\mu_W)+C_{6R}^{c}(\mu_W)\right)  
 -0.0170 C_{3R}^{d}(\mu_W) 
 +0.0880  C_{4R}^{d}(\mu_W) \nonumber \\
&+&0.0147 \left( C_{3R}^{u}(\mu_W) + C_{3R}^{c}(\mu_W) \right)
 -0.1732 \left( C_{4R}^{u}(\mu_W) + C_{4R}^{c}(\mu_W) \right)\nonumber \\
C_{8R}^{eff}(\mu_b) &=& 0.7032 C_{8R}(\mu_W) \nonumber \\
&+&0.1893  \left( C_{6R}^{d}(\mu_W)+C_{6R}^{s}(\mu_W)\right) 
 + 0.3670 C_{6R}^{b}(\mu_W) \nonumber \\
&+&0.1893 \left( C_{6R}^{u}(\mu_W)+C_{6R}^{c}(\mu_W)\right) \nonumber \\
&-&0.0894 \left( C_{3R}^{d}(\mu_W) + C_{3R}^{u}(\mu_W) 
        + C_{3R}^{c}(\mu_W) \right) \nonumber \\
&-&0.0801 \left( C_{4R}^{d}(\mu_W) + C_{4R}^{u}(\mu_W) 
   + C_{4R}^{c}(\mu_W) \right) 
\end{eqnarray}
where we have used $C_{7}^{eff}(\mu_W)=C_{7}(\mu_W)-C_{6}^b(\mu_W)$ 
and $C_{8}^{eff}(\mu_W)=C_{8}(\mu_W)$ on the right-hand sides of the 
above equations.

\renewcommand{\arraystretch}{1}
\begin{table}
\begin{tabular}{|c|c|c|c|c|c|c|c|c|c|c|}
$i$   &           1&           2&          3&            4&            5&          6&      7&      8&      9&     10\\ 
	\hline
$a_i$ &$\frac{14}{23}$&$\frac{16}{23}$&$\frac{6}{23}$&$-\frac{12}{23}$&$-\frac{24}{13}$&$\f{3}{23}$& 0.4086&-0.4230&-0.8994& 0.1456\\
\hline
$h_i$ &      2.2996&     -1.0880&$-\f{3}{7}$&$-\f{1}{14}$ &           0 &          0&-0.6494&-0.0380&-0.0185&-0.0057\\
$\bar{h}_i$& 0.8623&           0&          0&            0&           0 &          0&-0.9135& 0.0873&-0.0571& 0.0209\\
\hline
$r^1_i$ &   -0.1636&      0.3413&          0&            0&      -0.1242&          0&-0.1140&-0.0141& 0.0734& 0.0013\\
$r^2_i$ &   -0.5847&      0.7413&          0&            0&      -0.1032&          0&-0.1140&-0.0141& 0.0734& 0.0013\\
$r^3_i$ &   -0.1636&      0.0413&          0&            0&       0.1758&          0&-0.1140&-0.0141& 0.0734& 0.0013\\
$r^4_i$ &    1.9233&     -1.5327&     0.1714&      -0.1429&            0&          0&-0.4714& 0.0508& 0.0094&-0.0081\\
$r^5_i$ &    2.2996&     -1.9023&     0.1714&       0.1429&            0&          0&-0.6494&-0.0380&-0.0185&-0.0057\\
$r^6_i$ &    1.9233&     -1.1469&    -0.4286&       0.0714&            0&          0&-0.4714& 0.0508& 0.0094&-0.0081\\
$r^7_i$ &    2.2996&     -1.0880&    -0.4286&      -0.0714&            0&          0&-0.6494&-0.0380&-0.0185&-0.0057\\
\hline
$\bar{r}^1_i$&-0.0613&         0&          0&      -0.0316&          0&        0&-0.1604&-0.0325& 0.2258&-0.0049\\
$\bar{r}^2_i$&-0.2192&         0&          0&       0.1263&          0&        0&-0.1604&-0.0325& 0.2258&-0.0049\\
$\bar{r}^3_i$&-0.0613&         0&          0&      -0.0316&          0&        0&-0.1604&-0.0325& 0.2258&-0.0049\\
$\bar{r}^4_i$& 0.7212&         0&          0&            0&          0&        0&-0.6631&-0.1168& 0.0290& 0.0296\\
$\bar{r}^5_i$& 0.8623&         0&          0&            0&          0&        0&-0.9135& 0.0873&-0.0571& 0.0209\\
$\bar{r}^6_i$& 0.7212&         0&          0&            0&          0&        0&-0.6631&-0.1168& 0.0290& 0.0296\\
$\bar{r}^7_i$& 0.8623&         0&          0&            0&          0&        0&-0.9135& 0.0873&-0.0571& 0.0209\\
\end{tabular}

\medskip
\caption{The magic numbers with R--parity violation}
\end{table}
        
\begin{table}
\begin{tabular}{|c|c|c|c|c|}
$x$   &   $P_\gamma(x)$ & $P_g(x)$ & $P'_\gamma(x)$ & $P'_g(x)$ \\ \hline
$\left({170\over 100}\right)^2$ & 0.197 & 0.121 & 0.157 & -0.0558  \\
              1                 & 5/14  & 1/4   & 3/10  &   0      \\
$\left({170\over 200}\right)^2$ & 0.414 & 0.301 & 0.354 & 0.0378   \\
$\left({170\over 300}\right)^2$ & 0.560 & 0.445 & 0.499 & 0.176    \\
              0                 &   1   &    1  &    1  &      1  \\
\end{tabular}

\medskip
\caption{Values for the loop functions defined in the main text for several $x$'s.}
\end{table}
%

%


\begin{figure}
\begin{center}
\hbox to
\textwidth{\hss\epsfig{file=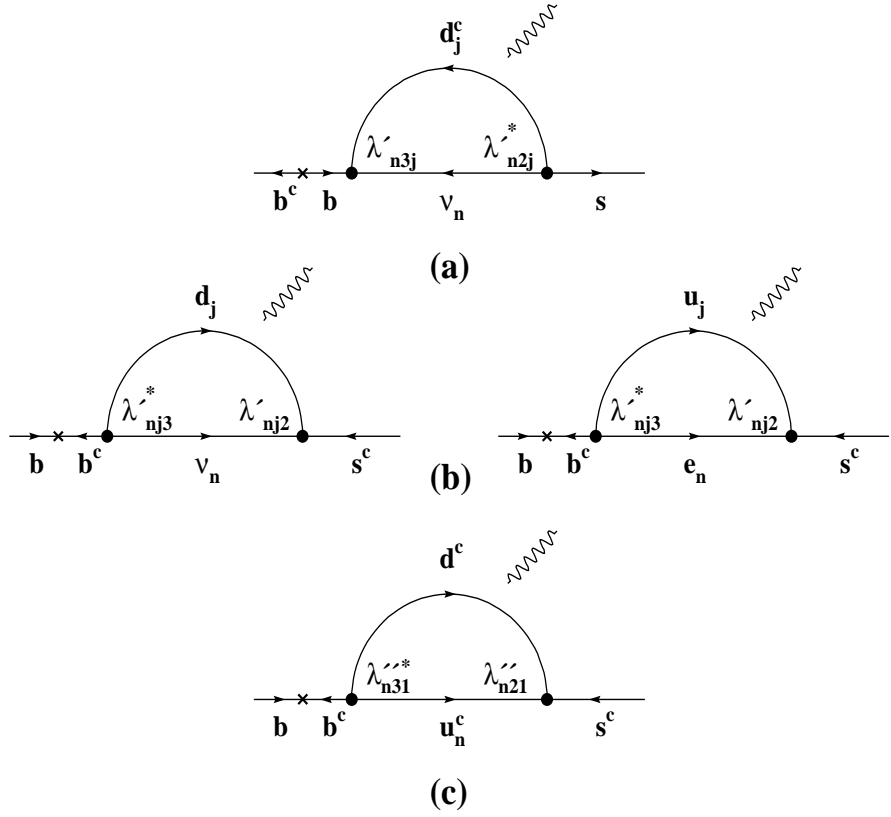,width=13cm,height=21.0cm}\hss}
\end{center}
\vskip -1.cm
\caption{ One-loop diagrams generating dipole moment interactions
for the $b\to s \gamma$ decay: the left-handed (a) and right-handed
type (b) from the lepton number violating couplings, and
the right-handed type (c) from 
the baryon number violating couplings.
}
\label{fig:bsg_rpv}
\end{figure}
\begin{figure}
\begin{center}
\hbox to\textwidth{\hss\epsfig{file=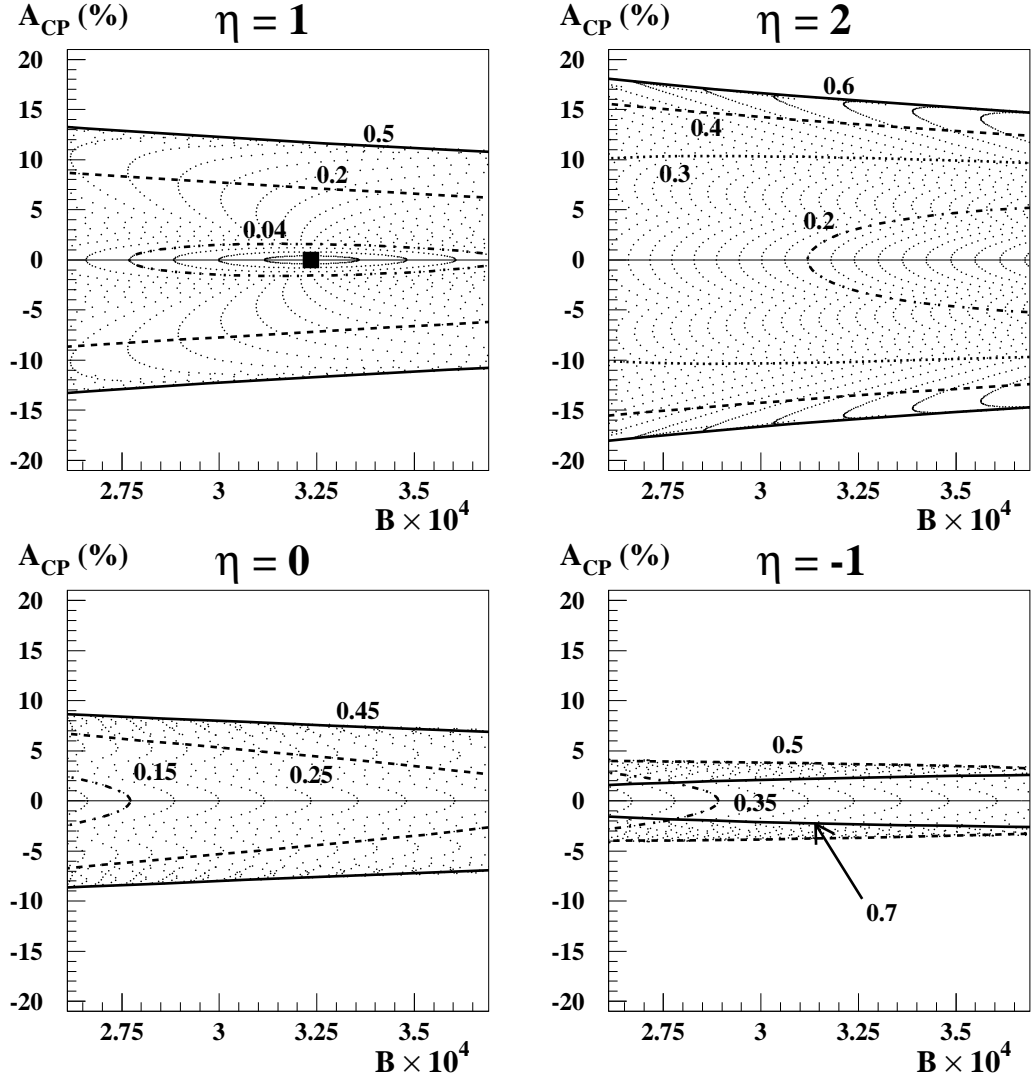,width=16cm,height=16cm}\hss}
\end{center}
\vskip -1.cm
\caption{ Scattered plots for $A_{CP}$ as a function of
B($B\rightarrow X_s\gamma) \times 10^4$ for the corresponding values of
$\eta$.  The SM prediction is marked as a filled square. 
The numbers are values of 
$|C_{6L}^b|$ for the corresponding contour lines.
The values of parameters of each case are shown in the text.
}
\label{fig:acp}
\end{figure}
%


\end{document}